# Dangers of Bayesian analyses and how to address them

Sander Greenland (corresponding author), Professor Emeritus,

Department of Epidemiology and Department of Statistics,

University of California, Los Angeles, California, U.S.A.

lesdomes@ucla.edu

Jason L. Oke, Departmental Lecturer in Medical Statistics,

Centre for Evidence-Based Medicine, University of Oxford, England

14 November 2026

**Abstract**. In light of the US FDA announcement supporting the use of Bayesian methods in clinical trials, we present a nontechnical review of problems with Bayesian analyses and methods to address them. Our focus is on the well-known sensitivities of Bayesian results to biases embedded in their prior distributions. This vulnerability calls for special safety features, including detailed justification of informative priors based on empirical-research literature rather than on singular expert opinions. Crucially, the information contributed by priors should be realistic in light of actual background data rather than merely judged so by available experts. We illustrate our recommendations with a textbook clinical-trial example which covered these points.

We then use the example to describe basic diagnostic procedures and safety recommendations for Bayesian analyses and their presentation. These include reference analyses which exclude informative priors, and justification for informative priors based on empirical-research literature rather than on philosophical arguments. We also critique some common priors that violate our realism requirements. One strategy to aid judgments about priors is to translate them into data to be added to the study data. The source of these prior data should be examined with the same critical attitude as the source of study data. We illustrate some simple summary methods for comparing prior and data information, including prior predictions of data, variance ratios, and their translation to effective sample sizes (ESS). Finally, we argue that Bayesian analyses should be treated as extensions or supplements to conventional analyses, rather than replace them, for that treatment puts prior construction and justification at the fore of the transition from frequentist to Bayesian analyses in teaching and practice.

**Corresponding author**: Sander Greenland

**Ethics approval and consent to participate:** Patients and members of the public were not involved in the design, conduct, or reporting of this methodological research. Only published tabular data are shown.

**Clinical trial number**: Not applicable.

**Consent for publication**: Not applicable.

**Availability of data and materials**: All data used are shown in the article tables.

**Competing Interests**: The authors declare that they have no competing interests.

**Funding**: There was no funding for this work.

**Authors' contributions**: Both authors contributed to the drafting and revision of the manuscript, and approved the final manuscript.

**Acknowledgements**: We are grateful to Krysztof Burdzy, Robert Matthews, Maggie Qian, David Spiegelhalter, Hector Velasco-Mondragon, Vera Wilde, and the referees for valuable comments on earlier drafts. The views expressed remain the sole responsibility of the present authors.

**Background**

The US Food and Drug Administration (FDA) recently declared itself to be open to Bayesian statistics for primary inference in clinical trials to support the efficacy and safety of drugs, and provided an extensive draft guidance document for that purpose (US FDA 2026). The change was described by some as a "a leap forward" beyond conventional frequentist data analysis, allowing lower drug development costs and reduced time for getting new treatments to market. These developments were covered in editorials and general news reports (e.g., Doshi 2026, Talpos 2026), and were greeted positively by some statisticians (Lee et al, 2026, Gelman et al, 2026a). Others, however were more critical (Evans et al 2026), raising cautionary warnings that Bayesian analyses share fundamental problems with frequentist analyses and add problems of their own via their hallmark feature: The introduction of a prior distribution (prior) for parameters.

Bayesian methodology has been promoted to remedy the long-noted problems of conventional statistical methods, and is now widely accepted in publications. There is limited empirical evidence that, in the hands of typical researchers, Bayesian methods lead to better inferences and decisions in real applications than - conventional frequentist methods. Our main concern however is that Bayesian prior distributions can introduce information into the analysis that might largely reflect misimpressions of literature, investigator biases, or unjustifiable certainty about effect sizes (Spiegelhalter et al. 2004, sec. 5.3; see also Spiegelhalter et al. 1994 and its discussion).

Another concern dealt with elsewhere is that priors can introduce simplifying assumptions (such as prior independencies) that conflict with actual background information in ways that go unnoticed by both statisticians and researchers, yet distort inferences (Greenland 2001, 2010). While the data models used by both frequentists and Bayesians alike are subject to similar objections to their validity, we will describe how priors for model parameters can open the door to direct distortions unlike those found in conventional data models.

Like many statisticians (e.g., Senn 2011) we believe that no single statistical methodology is free of limitations, and that Bayesian methods can be useful and are here to stay alongside frequentist methods. Researcher competency in understanding and reviewing Bayesian analyses has thus become essential, but requires a clear understanding of the similarities, differences, strengths, and limitations of conventional frequentist and Bayesian methods. Achieving widespread competency will further require that, *starting in basic statistics*,

Bayesian methods be taught in tandem with frequentist methods with emphasis on good practices and correct interpretations for both.

As with frequentist methods, Bayesian methods have an extensive and long-standing literature covering good practices for applications and reporting (e.g., Box and Tiao 1973; Box 1980; Spiegelhalter et al. 1994, 2004; Gelman et al. 2013; Kruschke 2021; McElreath 2020; van de Schoot et al. 2021; Held 2023; Gelman et al. 2026b). Yet, in a recent review of 49 phase III clinical trials using Bayesian methods, a third provided no information on the priors used in the analysis, and only 22% provided a justification of their priors (Ferreira et al, 2020); other surveys have noted similarly dismal reporting of priors (Kruschke 2021). This suggests that current medical reviewers and editors are unfamiliar with good practices and adequate reporting for Bayesian analyses .

We thus provide a nontechnical review aimed at clinical researchers, to provide journals with safety requirements for reports of Bayesian analyses. We focus on critical analysis and presentation of prior distributions, leaving aside technically more involved issues such as Bayesian trial design and controversies surrounding use of Bayes factors.

The requirements we would impose are that all informative priors (including ones labelled "weakly informative")

1. must be presented in the statistical methods section of the primary paper in enough detail to allow reproduction by the reader;
2. must be given a full justification in relation to background topic information with citation to supporting empirical literature (if extensive this might be in an appendix, but is essential);
3. must be presented with a comparison of the information that the data and the prior contributed to estimates of the targeted effect.

In most medical applications, requirement 3 can be met by simple practices such as graphical comparisons of prior and data information (e.g. see Higgins & Spiegelhalter 2002; Spiegelhalter et al. 2004 Fig. 3.7; Greenland 2007; van de Schoot et al. 2021; Modin et al. 2025). Space considerations may however preclude graphs in papers; hence we will illustrate some simple diagnostics for examining prior distributions and their correspondence with accepted facts.

Finally, we focus exclusively on issues in basic teaching and in preparing Bayesian analyses for publications. We argue that Bayesian analyses should be treated as extensions or supplements to conventional analyses, rather than replace them, for that treatment puts prior construction and justification at the fore of the transition from frequentist to Bayesian analyses in teaching and practice. For Bayesian methods in trial design, conduct, and decision-making see for example Spiegelhalter et al. (2004).

## Differences and parallels of frequentist and Bayesian methods

### *Frequentist vs. Bayesian methodologies*

To evaluate the issues, one should understand how Bayesian analyses extend and differ from the frequentist analyses that they are proposed to replace. Appendix 1 provides an overview of how we would approach elementary coverage of fundamental Bayesian principles while minimizing mathematical issues.

With those principles in place, it should be emphasized that the division into “frequentist” and “Bayesian” analyses is crude, as there is a huge variety of philosophies and methods that fall under each label (e.g., it was once estimated that over 46,000 logically distinct approaches could be labelled as “Bayesian” (Good, 1971)). Even with simplification one can discern several distinct and often conflicting schools within both frequentism and Bayesianism (Goodman 1993; Spiegelhalter et al. 2004 sec. 3.20).

Thus, in practice “frequentist” and “Bayesian” only indicate whether statistical quantities are defined exclusively in frequentist terms of probabilities of observations given hypotheses (e.g., P-values, confidence levels, likelihood ratios), or instead if results are phrased in Bayesian terms of probabilities of hypotheses given observations (e.g., prior and posterior probabilities and intervals).

### *Conventional frequentist practices and Bayesian alternatives*

Conventional frequentist analyses produce P-values for statistical hypotheses which are then used to claim or deny “statistical significance” of results, usually based on whether $p \leq 0.05$ or $p > 0.05$ for the null hypothesis of “no treatment effect”. They also produce interval estimates for effect sizes labelled as “confidence intervals” (CI) which are supposed to measure the statistical precision of the results. These quantities are derived from assumptions about data generation that assume any influences not accounted for can be treated as “random” errors. Throughout most of the 20$^{th}$ century these statistics and interpretations dominated research, where they fell prey to numerous misinterpretations (25 of which are listed in Greenland et al. 2016a) in both research reports and statistical primers.

Common among them is mistaking frequentist quantities for Bayesian statements. This often involves mistaking a P-value for a hypothesis probability, as when the P-value for "no effect" is misdescribed as "the probability that chance alone produced the association": "Produced by chance alone" is simply another way of stating the hypothesis that there was no effect; more precisely, it is a positively stated version of the null hypothesis that the observed association was *not* caused by treatment or bias.

Another common example is mistaking the 95% attached to a conventional confidence interval (CI) for the probability that the true effect falls inside the reported interval. In reality, the interval only displays all the effect sizes that would have $p>0.05$ under the assumptions that were used to compute the interval, and that when the assumptions are correct, intervals computed under those assumptions (not only the one we observed) will contain the true value 95% of the time (Greenland 2025). Neither of these frequentist interpretations say there is 95% probability that the true effect is inside the one interval we observed.

The statement that the true effect is inside our observed interval is called an interval hypothesis; thus the probability the effect is inside the interval is a hypothesis probability, not a frequency. Unlike conventional frequentist methods, Bayesian methods *do* provide probabilities for hypotheses, called prior probabilities if formulated before seeing the analysis data, and posterior probabilities if formulated after seeing the data. The posterior probabilities are computed from prior and data probabilities via Bayes' theorem using the same assumptions as frequentists use for the data probabilities (see Appendix 1), hence are described as "Bayesian".

A key question is whether these probabilities represent anything more than quantified opinions or personal bets. Some Bayesians say that all probabilities are merely such bets, and that the Bayesian format is more rational and honest in handling this limitation of human inferences (de Finetti 2017). Not all Bayesians agree, noting that Bayesian data probabilities are based on the same assumptions about data generation as used in frequentist analyses, and should be checked against data frequencies (Box 1980).

Some advocates for Bayesian methods take the common misinterpretation of frequentist results as if they were Bayesian to show that researchers want and need Bayesian results. They further argue that correct frequentist inference is hopelessly complex and incoherent in a way remedied by Bayesian statistics, and that technically correct description of Bayesian interval estimates is straightforward and intuitive by comparison (Lindley 1975; McElreath 2020). For example, the "95%" in a 95% Bayesian interval is supposed to represent the credibility or betting probability the presenter would assign to the interval, as in "I would bet 95% that the true value is in this interval"; this could be a bet before (prior to) or after (posterior to) seeing the analysis data.

*Frequentist defences and reforms*

Defences of frequentist methods sometimes respond that correct frequentist interpretations are less ambitious than Bayesian ones, and that Bayesian misinterpretations merely show that users are trying to draw stronger conclusions from statistical analyses than are warranted. A major concern is that, *even when pre-specified*, priors add another avenue for guiding results toward desired conclusions, beyond study problems such as selective recruitment and contrived treatment or outcome definitions. Then too, for better or worse, conventional frequentist methods are easy to apply automatically, yet the resulting estimates will be accurate to the extent that assumptions about data generation (e.g., randomization) are correct; in contrast, Bayesian priors add a layer of assumption sensitivity that requires more background information to check. (Efron, 1986).

Additionally, it has long been argued that misinterpretations of conventional frequentist results can and should be addressed by simple reforms to teaching, practice, and presentation. Among reforms now widely adopted are to present P-values directly as continuous quantities (e.g., Lehmann 1986 p. 70-71) rather than describe them in reference to a dichotomy based on a cutpoint such as 0.05, and to always provide and discuss an interval estimate for the effect (Yates 1951, Rothman 1978, Rothman 1986).

Semantic arguments further note that the term "statistical significance" invites confusion with practical significance, while the term "confidence interval" invites confusion with a Bayesian interval (Rafi & Greenland 2020; Greenland et al. 2022). As a consequence, more neutral terms like "statistical compatibility" and "compatibility interval" have been promoted as replacements; a P-value p then becomes an index of compatibility, consonance, or goodness of fit between the data and a hypothesis or model, while the interval becomes the range of effect sizes that have $p>0.05$ and thus have "reasonable" compatibility with the data according to the conventional 0.05 cutoff (its abbreviation remains CI, for "compatibility interval" or "consonance interval"). (Amrhein & Greenland 2022, Berner & Amrhein 2022,

Greenland et al. 2022, Rafi & Greenland 2020, McShane et al. 2024, Greenland 2025 sec. 3.1, Rovetta et al. 2025a,2025b)

*Insights from numerical differences*

What if we look beyond differences and problems in interpretations to the purely numerical values of statistical results? We then find that the key distinction between conventional frequentist and Bayesian results is that the latter reflect information added in the form of prior distributions on model parameters (Gelman & Robert 2013; Gelman et al. 2017).

In most medical applications, the degree to which Bayesian intervals and one-sided (directional) hypothesis probabilities differ from conventional frequentist interval estimates and one-sided P-values depends on the amount of information in the prior compared to the amount of information in the data under the assumptions about data generation. (Casella & Berger 1987a; Greenland & Poole, 2013a) In those cases, the frequentist results could still be made to numerically approximate Bayesian results by injecting the prior into the frequentist analysis as part of the data-generating (sampling) model.[1] Nonetheless, conventional frequentist analyses do not do this, and thus their resulting estimates can exhibit large numerical differences from Bayesian estimates.

Conversely, in the world of clinical trials there are priors that have no realistic Bayesian justification in terms of background medical data. Examples have spikes of probability on the null hypothesis of no effect (which have been used for computing Bayes factors and "Bayesian significance tests"): In the vast majority of medical research there is no real data that could justify such extreme concentration of prior probability. Consequently, null spikes can lead to excessive certainty in claims of "no effect" (Casella & Berger 1987a, Casella & Berger 1987b, Greenland 2009b, Greenland 2011, Gelman 2013, Greenland and Poole, 2013a 2013b), thus duplicating in Bayesian terms the frequentist fallacy of treating statistical nonsignificance as indicative of no effect (Greenland 2011, 2017, Greenland et al. 2016a).

In sum, despite the fact that we can build numeric parallels between frequentist and Bayesian methods, the two approaches can diverge widely in practice. This divergence is largely traceable to priors, which is why Bayesian analyses need to describe and critically analyse priors with the same attention to detail as given to data description and analysis.

[1] This can be done by entering the prior distribution into the frequentist model as a known parameter generator, as in random-effects or hierarchical (multilevel) models with known effect distributions (as in Greenland 2000), or by entering the log prior as a penalty function on the loglikelihood (as in Greenland 2009a and Cole et al. 2014).

**An illustrative example**

To illustrate basic concepts and good practices, Spiegelhalter et al. (2004) used the data in Table 1 from the GREAT trial (GREAT Group, 1992), which randomised patients who had a recent myocardial infarction to treatment with the thrombolytic treatment anistreplase administered either in home or in hospital. The estimated risk ratio (RR) was (13/163)/(23/148) = 0.513. From a conventional frequentist analysis[2], the estimated variance for the log risk ratio was 0.1075, and the 95% compatibility ("confidence") limits for RR were 0.270 to 0.976, with a (one-sided) P-value of p=0.019 for the hypothesis that RR≥1 (no advantage and possible disadvantage for home treatment).

*Table 1. Data on 3-month mortality in the Grampian Region Early Anistreplase Trial (GREAT group 1992).*

| | Treatment location | |
|---|---|---|
| | Home | Hospital |
| Deaths | 13 | 23 |
| No. patients | 163 | 148 |

Given the wide interval estimate, however, the trial could not reliably discern slight (2%) from dramatic risk reduction (73%) in mortality at home. Spiegelhalter et al. (2004, ex. 3.6) thus showed how to reduce uncertainty by adding in cautious prior belief about the likely magnitude of effect, as elicited from a senior cardiologist. This expert judged that treatment given shortly after a myocardial infarction would most likely reduce mortality by 15-20%, with no reduction and a reduction as large as 40% both being unlikely.

We will deviate slightly from Spiegelhalter et al.'s example (which used odds ratios) and interpret this judgement as a 95% prior-probability interval for the risk ratio RR of 0.60 to 1.00. From that we impute a lognormal prior distribution for RR with a variance for the natural log of RR of 0.0170 and a median for the RR of 0.7746 (a risk reduction of about

[2] Computed using the Wald approximations in Rothman et al. (2008, Ch. 14) for which the variance of ln(RR) is 1/13+1/23−1/163−1/148.

23%)[3] which produces a prior probability for RR≥1 of 0.025. Following Spiegelhalter et al. we will not use prior information about the background risk, so technically our analysis will be partial Bayes or semi-Bayes (Cox 1975, Greenland 2006).

There are many ways to compute the Bayesian posterior distribution from this prior information and the data. Assuming approximate normality, we may use inverse-variance (information) weighted averaging of the prior mean and the conventional estimate for the log RR (Box & Tiao 1973, sec. 1.2.3; Spiegelhalter et al. 2004, sec. 3.7; Greenland 2006; Rothman et al. 2008 Ch. 18). This produces a posterior distribution for RR that is lognormal with a median of 0.73, a 95% interval of 0.58 and 0.93, and a posterior probability for RR≥1 of 0.005.[4] While these results agree qualitatively with the conventional results, they point to a much smaller benefit of home treatment, as seen in other trials of pre-hospital thrombolytic administration conducted at the time of the GREAT trial, several of which influenced the expert's prior (Morrison et al. 2000, Table 3; Spiegelhalter et al. 2004, p. 72).

Nonetheless, before accepting such numbers as "the" Bayesian result from the study, we should ask penetrating questions such as: Would other experts in the field agree with the opinion on which the prior was based? If we used their priors, would we get a different conclusion? And are the cardiologists' beliefs really distributed lognormally? Those questions should give us pause about using the Bayesian results from this prior as if they were definitive findings (Spiegelhalter et al. 2004, sec.5.3).

We should also question the sense in which the Bayesian results represent a "data analysis". Foremost, one should not mistake Bayesian results for descriptions of what was observed: Not only is the prior interval more precise than the conventional interval, but the observed RR of 0.513 was not even inside the 95% prior interval – in fact the prior probability that RR≤0.513 was 0.0008 or 0.08%!

**Comparing frequentist and Bayesian analyses**

One can always compute and compare frequentist and Bayesian analyses of the same data, and that comparison corresponds to our third reporting requirement. When we use the same data model for both analyses, disagreement in the results must arise from the information added by the prior distribution. It is thus imperative to compare the prior and data information, and demonstrate that the prior information does not overtly conflict with the observations, and is at least as valid and reliable as the data information (Spiegelhalter et al. 2004; Qian 2026). Figure 1 shows the information from the data and the expert prior in the GREAT example, along with their merged information in the posterior, where the information is expressed in the form of probability curves[5] for the log risk ratio; for ease of substantive interpretation the horizontal axis labels are for the risk ratio. It can be seen that the expert prior is more concentrated than the data curve and thus contains more information than the data supplies; hence the posterior probability curve is much closer to the prior probability curve than to the data probability curve.

---

[3] The prior RR median was taken to be the geometric mean of the 95% prior limits, $(0.60*1.00)^{½}$ = 0.7746, with the ln(RR) variance calculated from its interval width as $(\ln(1.00/0.60)/(2\cdot 1.96))^2 = 0.0170$.

[4] From a posterior ln(RR) variance of $v = 1/(1/.1075+1/.0170)$, mean $v\cdot(\ln(.513)/.1075+\ln(.775)/.0170)$.

*<Fig. 1. GREAT example: Probability curves for the log risk ratio from the expert prior, the data (see footnote 4), and the posterior curve from merging the two using Bayes theorem; for ease of substantive interpretation the horizontal axis labels are for the risk ratio.>*

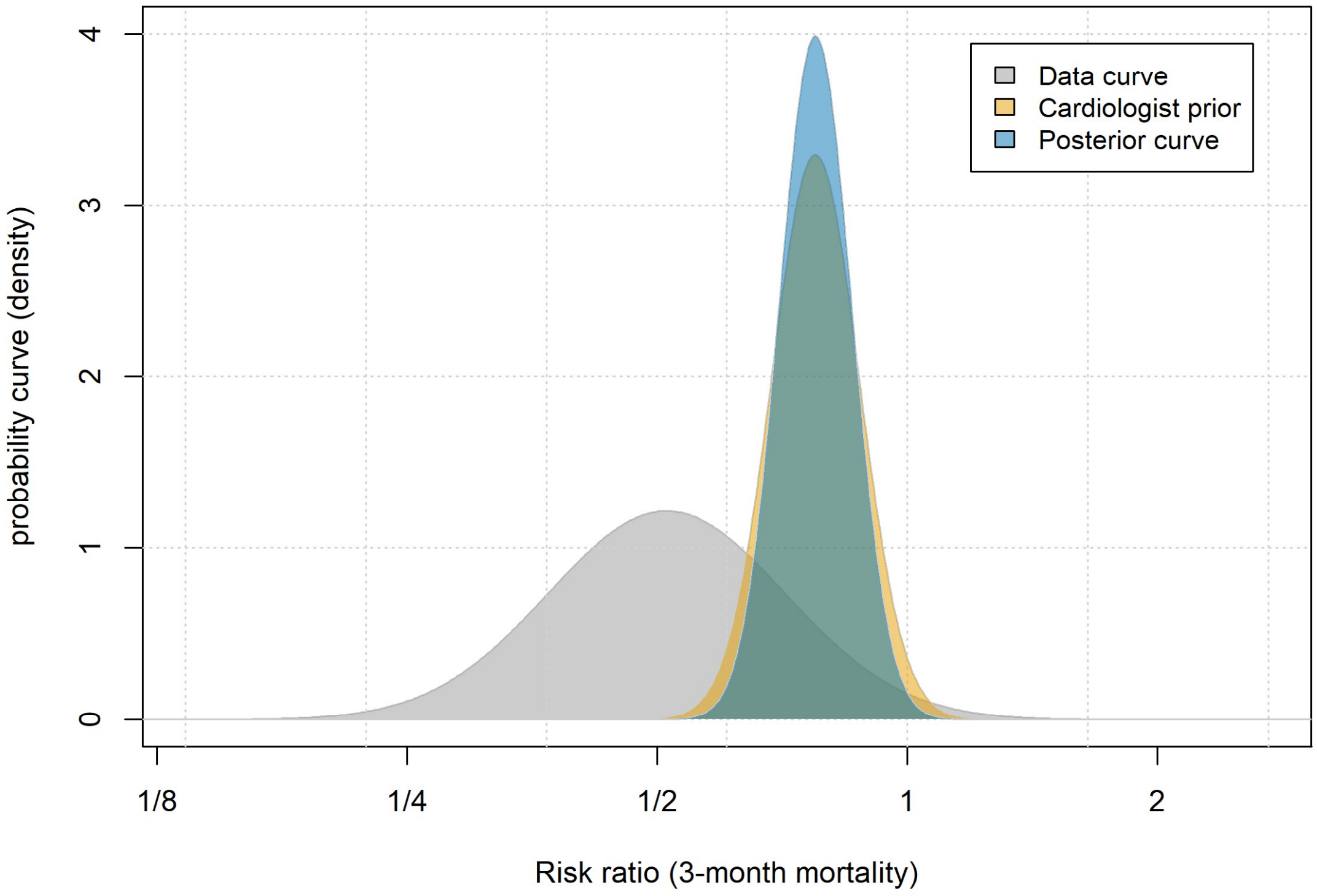

[5] Prior and posterior densities plus the normal(ln(0.513),0.1075) approximation to the likelihood function, which is also the estimated sampling distribution for the estimated log risk ratio.

A quantitative way to assess the compatibility or conflict between the prior and the data is to treat the prior mean and variance as if they were frequentist estimates from another study whose data appear as described in Appendix 2. We can then use conventional methods to compare those estimates to the estimates from the current data (Greenland 2006; Rothman et al. 2008 Ch. 18; Held 2023). In the GREAT example this process yields p=0.24 for comparison of the prior RR of 0.7746 to the GREAT frequentist RR of 0.513.[6] The high compatibility of these estimates can also be seen from the fact that the prior 95% interval (0.60,1.00) is almost entirely contained inside the frequentist interval (0.27,0.98). This containment further tells us that the prior is more precise than the frequentist result; thus the posterior interval (0.58,0.93) is based more on the prior than the data, which is all the more reason to scrutinize the prior closely.

**Types of priors and their problems**

*Expert opinions*

Priors are sometimes built from solicitation of expert opinions or bets labelled as "expert knowledge" or "prior beliefs". While the information introduced this way may be valuable, its connection to real studies may be unclear and may even be distorted by statistical misinterpretations (such as those listed in Greenland et al. 2016a) or by publication biases such as the tendency for initial reports to display "significant" P-values (p<0.05) and overestimated effects (van Zwet et al. 2021). The analyst and reader thus may not know whether expert opinions are trustworthy unless they themselves are also experts in both the relevant literature and in its statistical problems – in which case they could use their own knowledge to formulate a prior they would trust. At the very least, we would want "expert" prior intervals to encompass all possibilities considered reasonable by all experts in the topic area.

Even if it is decided to rely on expert opinions, there are practical problems with eliciting and translating those into a form that can be combined with the data. One challenge is that such prior information tends to be vague. For example, we may know that benefits will increase as dose increases up to a plateau whose location is unknown and varies across patients, and then decline as toxic effects accrue. A major challenge for prior construction is to map these types of vague qualitative descriptions into precise continuous probability distributions so that they can be used in Bayesian computations.

---

[6] P-value (2-sided) from the Z-score for the difference ln(0.775/0.513)/(0.0170+0.1075)$^{½}$.

Our largest concern may however be that priors derived from expert elicitation are particularly susceptible to confirmation bias from entrenched positions and to overconfidence as displayed by prior precision that cannot be justified from actual pre-existing literature. Such cognitive biases can be amplified by statistical fallacies (such as mistaking "nonsignificance" as support for the null hypothesis) and by conflicts of interest. Given that it has been difficult to detect and limit conflicts of interest among authors and sponsors of clinical guidelines (Lenzer, 2013), can we realistically hope to police the elicitation of priors? Evidence-based medicine was meant to de-emphasize "intuition, unsystematic clinical experience, and pathophysiologic rationale" in clinical decision making (Guyatt et al, 1992), yet expert priors open a door for biased opinions to affect statistical results in crucial ways.

In light of the above problems, we may expect different experts to produce very different priors. If such differences are missed because only one expert camp is consulted, the resulting priors may be seen by other experts as a source of bias in the results. If the differences are recognized, they can be accommodated by a very broad prior that encompasses the range of opinions, at the possible cost of reflecting no expert's opinion; a related strategy mixes together differing priors (Moatti et al, 2013).

*Default, skeptical, and reference priors*

To avoid the labor of prior construction (especially the minefield of expert elicitation), there have been proposals for priors that can be used as defaults for analysis. These default priors expose a difference between idealized philosophy and actual practice of Bayesian analysis: Rather than summarizing carefully collected external information, default priors may appear to be questionable devices to enable computations and statements of Bayesian form (Gelman et al. 2017).

Default priors are described as neutral or "objective", but some are extremely biased toward unjustified extreme skepticism about effects. Although skepticism may be necessary to avoid wasteful pursuit of dead ends, excessive skepticism can suppress new discoveries and obstruct study of important anomalies. We will return to this problem in our discussion of spiked priors. Moderate skepticism may however be justified by past experiences with promising treatments often failing to achieve initial optimistic expectations. Methods for creating skeptical priors that reflect actual prior information are described by Spiegelhalter et al. (2004, sec. 5.5.4) and an example will be given below.

At the opposite extreme from skepticism, priors can be nearly non-informative in the sense of injecting information into the analysis that would appear slight to all observers. There is considerable variation in the details of such *reference priors*; examples include priors with 95% limits that extend far beyond what anyone would expect for the true RR. Nonetheless, because those priors add little information, the posterior intervals they produce tend to be close numerically to frequentist interval estimates, and the posterior probabilities of one-sided hypotheses (such as noninferiority and superiority) they produce are close to P-values for those hypotheses (Casella & Berger 1987a, Greenland & Poole 2013a). Furthermore, those reference-posterior quantities can exhibit superior frequentist properties and match background expectations better than can conventional frequentist statistics (Firth 1993, Greenland and Mansournia 2015, Greenland et al. 2016b).

When the results from frequentist and reference-Bayesian analyses numerically coincide under the assumed data-generation model, their only difference can be in interpretation. For example, in frequentist analyses of simple tabular data such as Table 1, bias adjustments or "continuity corrections" are sometimes made by adding ½ to each interior count before calculations are done. The resulting CIs are close (and in some cases identical) to the posterior intervals from a Bayesian analysis using a particular reference prior for the log ratio which assigns 95% prior limits of 1/648 and 648 to the RR (Greenland and Mansournia 2015); see Appendix 2. In the GREAT example, adding ½ to each cell results in an approximate 95% frequentist CI of (0.28,0.98), which to two digits equals the posterior interval from the reference prior. The practical difference is therefore only in the interpretation: for a frequentist, (0.28,0.98) shows all the RR that have $p>0.05$ after the adjustment; whereas for a Bayesian, one should bet 95% that the true effect is inside (0.28,0.98) *if* the prior information about the RR is as little as reflected in the reference prior.

An objection to reference-Bayes analysis is its assumption that there is almost no information about the effect beyond that in the data. This assumption is patently absurd in practice, because no human trial would be initiated without some suggestion of a net beneficial effect from earlier studies, which would also provide some rough indication of the effect size to expect. Conventional frequentist analyses and analyses with reference priors should thus incur contextual objections if used alone to make decisions about effect size or direction. Nonetheless, they remain useful for showing the information provided by the study alone under the assumed data model, thus serving as reference points to gauge the impact informative priors have on Bayesian posterior results.

*Priors as imagined data*

In examples like GREAT in which the frequentist estimator and the prior have approximately lognormal distributions,[7] Bayesian posterior intervals will be numerically close to what we would get from conducting a frequentist meta-analysis of the actual study and another hypothetical study whose frequentist results numerically match those computed from the prior (Greenland 2006; Rothman et al. 2008 Ch. 18). Table 2 shows data that represent the expert prior in the example as if it were derived from a frequentist analysis of prior data.[8] The table thus exhibits a point estimate of 0.77 and a 95% confidence ("compatibility") interval of (0.60,1.00), which equal the prior median and the 95% prior limits.

*Table 2. Data from a hypothetical trial of 3-month mortality constructed to produce point and interval estimates that closely match a lognormal prior for RR with median 0.77 and 95% limits of 0.60 and 1.00.*

| | Treatment location | |
|---|---|---|
| | Home | Hospital |
| Deaths | 118 | 118 |
| No. patients | 10,000 | 7,746 |

Figure 2 shows a forest plot of what a meta-analysis would look like combining the actual data in Table 1 with the prior data in Table 2. From the figure it is again apparent that the prior makes a much larger contribution to the posterior than does the GREAT data.

<*Figure 2. Forest plot for the frequentist meta-analytic analog of the Bayesian analysis with expert prior.*>

[7] A more precise technical condition is that the likelihood function and prior are both approximately normal in shape.

[8] Appendix 2 shows how this is done by making the death counts of equal size to reproduce the prior variance. The denominators are then set to make the table RR = 7,746/10,000 = 0.7746, the median of the expert prior, and also set large enough so that only the death counts contribute to the prior variance.

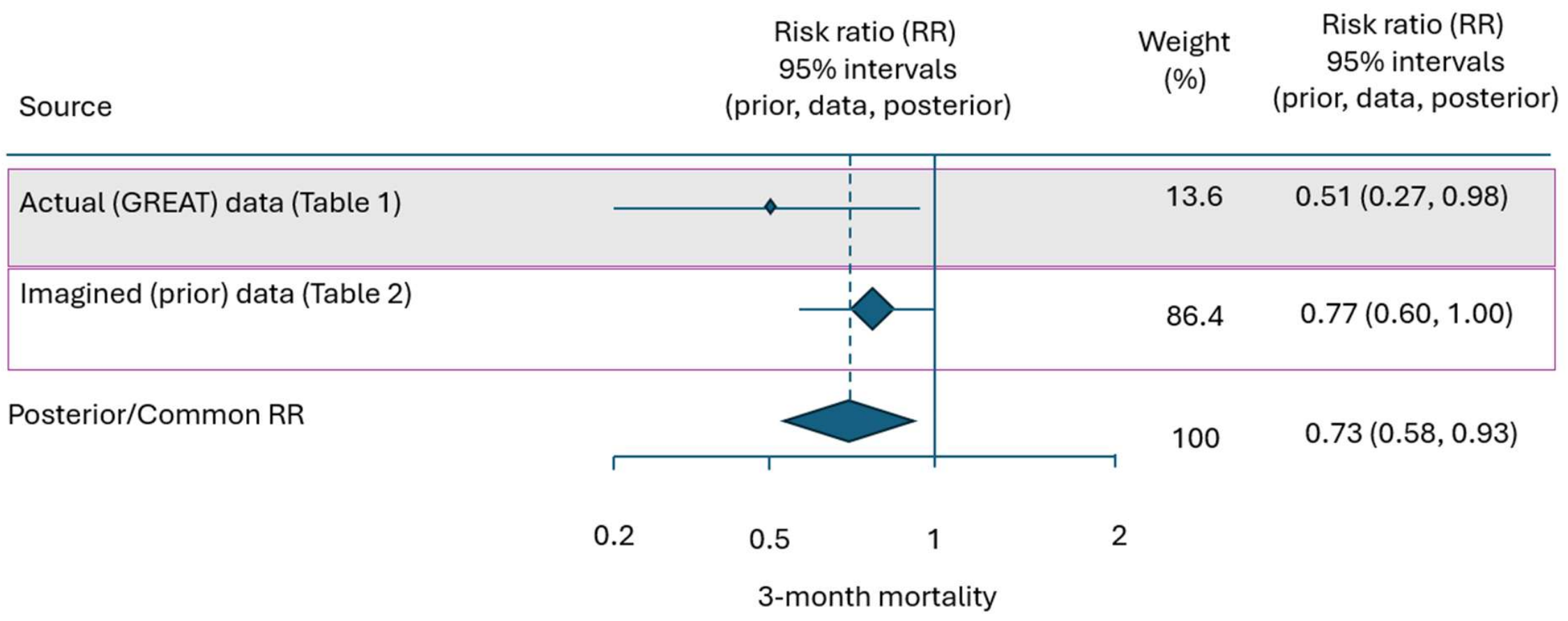

This figure suggests a cynical view of an expert prior: Suppose the analyst had not done a Bayesian analysis but instead first made up a fake trial, then added the fabricated data from it to the actual data, and finally reported estimates from a conventional frequentist meta-analysis on the combined data as if all the data were real. Such an act would have been decried as corrupting the actual data with fraudulent data. This was found to have been done in a trial of ivermectin and rightly called out (Hill et al. 2022). But, had the original authors instead translated their fake data into a Bayesian prior and used that in a meta-analysis, it might have instead held up as an example of Bayesian methodology in action.

The fabricated-trial analogy applies most strongly to priors that are undisclosed or unjustified, and an openly declared prespecified prior is quite different from fraudulent data (Qian, personal communication). Given such an auditable prior, its translation into hypothetical trial data can help evaluate whether it is supplying information beyond what could be justified from the actual background literature on the topic (Greenland 2006; Rothman et al. 2008 Ch. 18). For example, we would reject priors for which the prior-trial size is beyond the total from existing trials. Similarly, we would reject a prior whose shape could not be reproduced from the kind of real-data information available in the application context. More precisely, we would require that the prior can be translated into a frequentist data summary (such as a *likelihood function* or a *compatibility distribution*; Rafi & Greenland 2020) obtained by combining a series of plausible experimental results.

*Spiked (point-mass) priors and "Bayesian significance tests"*

Highly controversial examples of priors that dramatically fail realism requirements include priors with a spike or mass of probability (typically 50%) on the point null hypothesis of "no effect", as used in what are known as "Bayesian significance tests". Spiegelhalter et al. (2004, sec. 5.5.4) review historical arguments for using such priors. There are however Bayesian arguments against them. One is that they turn Bayesian inference into circular reasoning by partially assuming the null hypothesis in order to measure support for the null hypothesis, rather than by using documented background information to improve inference accuracy; they thus induce or aggravate confirmation bias (Greenland 2009b, Greenland & Poole 2013a, Gelman 2013, Greenland & Poole 2013b).

In effect, priors with spikes of mass on the null give a Bayesian form of the common frequentist fallacy that data support the null because the significance test failed to reject the null (Greenland 2011, 2017). For those like us who demand an empirical rationale for priors, spiked assume there are background studies that never exist for real medical research : A null spike of ½ probability asserts there is a 50% chance there is background studies that prove the null with absolute certainty; that is, we would place a 50:50 bet that there is decisive information that *logically* proves the null. Without such information (such as a point prediction by an extensively verified physical law), a probability spike at the null is an example of how initial probability assertions (priors) can misrepresent ignorance as strong background knowledge (Popper 1957), knowledge that would never exist in an ethical trial of a new treatment protocol.

*Priors formulated from external data*

Priors taken solely from pre-existing studies can be acceptable to the extent that those studies are seen as comparable or exchangeable with the current study and representative of such studies (Spiegelhalter et al. 2004, sec. 5.4). This exchangeability requirement parallels requirements for use of historical controls (Pocock 1976; Schmidli et al., 2014) and for combination of studies in meta-analyses (Higgins & Spiegelhalter 2002, Rothman et al. 2008, Ch. 33).

Consider a prior derived directly from a collection of pre-existing studies, such as the Cochrane Collaboration (van Zwet et al. 2021). The justification of the prior then largely hinges on the extent to which those study reports are reliable, valid, and come from a setting in which we expect about the same effect size after feasible adjustments are made for any relevant differences between the external studies and current study (Spiegelhalter et al. 2004, sec. 5.4). In that case, the studies are said to be *partially exchangeable* with the current study.

The simplest adjustment downweights the external studies relative to the current study, but there are many ways to do that and no conventions on which to use. When no downweighting is done, the resulting posterior intervals may be numerically the same as what one would have obtained from a frequentist meta-analysis, whereas when the downweighting becomes extreme the posterior interval approaches the conventional frequentist interval from the current study. More sophisticated analyses will include adjustments for differences between the external studies and the current study (Pocock 1976, Dahabreh et al. 2024).

**Prior sensitivity analysis**

As may be clear from the discussion so far, Bayesian analysis with informative priors must grapple with the fact that the number of defensible priors that one could use is almost unlimited. To address the inevitable uncertainties and potential controversy about which prior one should use, the analysis can be repeated over a range of priors, a process called prior-sensitivity analysis (Spiegelhalter et al. 2004; Gelman et al. 2013; McElreath 2020; van de Schoot 2021; FDA 2026).

A detailed example of this approach is the Bayesian re-analysis of the DANFLU-1 trial (Modin et al. 2025). We will provide a much simpler illustration using the GREAT example, where the expert prior had 95% limits of 0.60, 1.00 and a probability of harm from home application is only 2.5%. There we saw the expert posterior limits of 0.58, 0.93 agree with the conventional frequentist limits of 0.27, 0.98 in making home treatment look more effective than hospital treatment. But the posterior heavily reflects expert skepticism about both harms and large benefits of home treatment, with risk reductions of over 50% having only 0.08% probability. In contrast, using only the data information leaves a 70% reduction looking like a reasonable possibility.

Consider then a prior completely neutral or in equipoise about whether at-home or hospital is better, but which is skeptical of a large effect in either direction. In one such prior the risk ratio RR is lognormal with median 1 and 95% limits of ½ and 2 (see also Spiegelhalter et al. p. 71-72). This prior assigns 50% prior probability to both $RR \geq 1$ and $RR \leq 1$, and for any value r for RR, the prior assigns the same probability to $RR \geq 1/r$ and to $RR \leq r$, so is completely symmetric around the null value on the log scale. The corresponding posterior median RR is 0.70 with 95% limits of 0.44, 1.12. The posterior probability that home treatment was harmful ($RR \geq 1$) is 0.068 or about 7%, which seems less definitive about the superiority of home treatment than do the conventional results from no prior and the results from the expert prior. It also assigns 8% probability to $RR < ½$, more in accord with the conventional result.

An issue with sensitivity analyses is that there is no precise guidance on the variety of priors or data models to explore, and is vulnerable to manipulation to produce desired results (Qian 2026). This quandary is partially addressed by reverse-Bayes analyses, which finds priors that produce Bayesian results of special interest such as having RR=1 as the upper boundary of the 95% posterior interval and a 97.5% posterior probability for RR<1 (Matthews 2001, Matthews 2018, Held et al. 2021). In the GREAT example that can be done using a lognormal prior for RR with median 1 and 95% limits of 0.10 and 10, and variance of 1.381 on the log scale. This prior assigns 28% probability to RR<½, and results in posterior 95% limits of 0.29 and 1.00 with probability for RR≥1 of 0.05/2 = 0.025, and a 41% posterior probability of RR<½. These numbers are very close to their conventional frequentist analogues, showing that the prior needed to make the upper 95% posterior RR limit equal 1 is not very informative compared to the data, and that only mild skepticism about large effects would cast doubt on the decisiveness of the GREAT data for determining the direction of the effect.

Figure 3 displays the probability curves (densities) for the priors considered above along with the probability curve (approximate likelihood function) for the data, showing the large differences in information among them.

*<Fig. 3. GREAT example: Probability curves for 3 prior distributions and the data; the data curve is a normal(ln(.513),.1075) approximation to the likelihood function for log RR; for ease of substantive interpretation the horizontal axis labels are for RR.>*

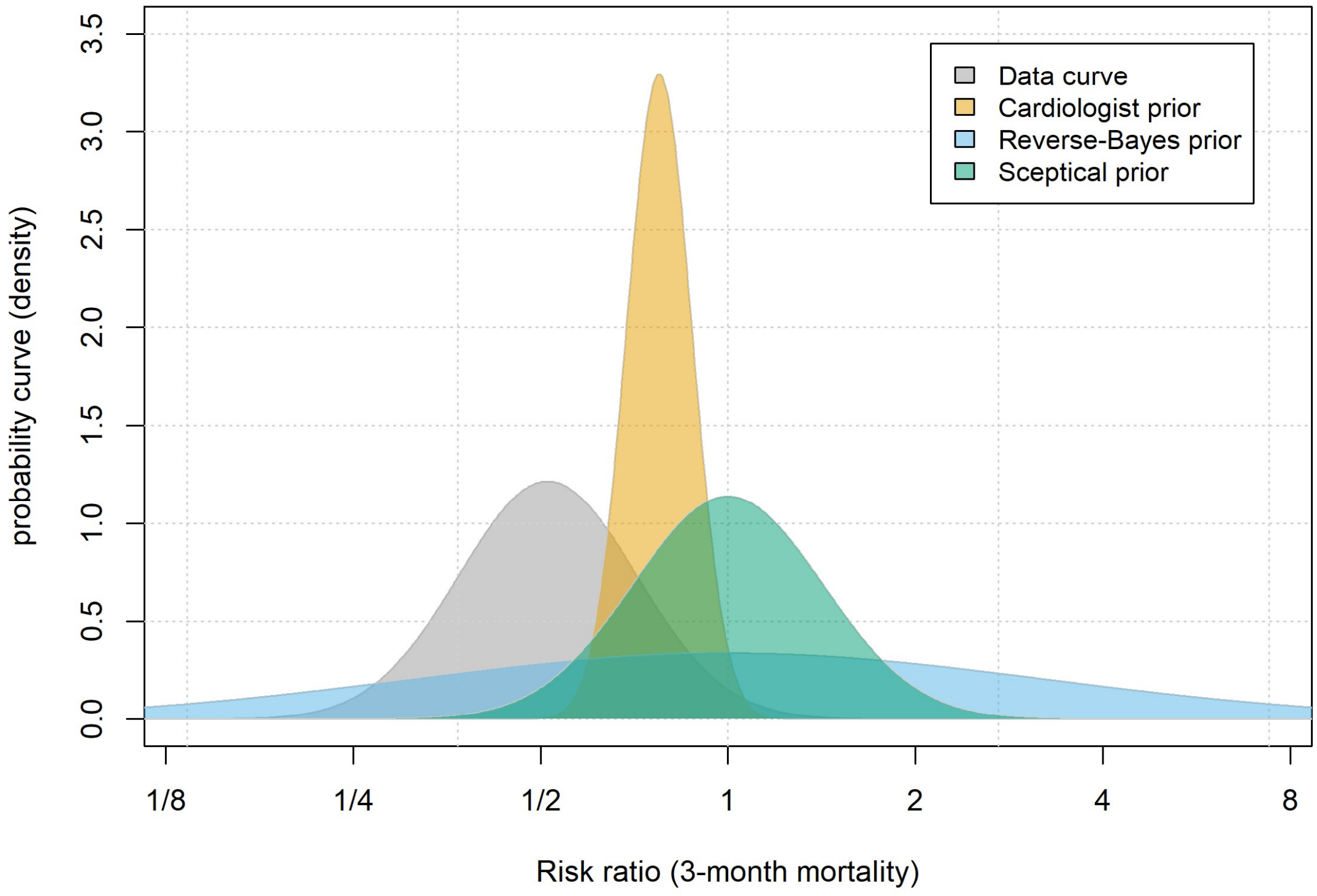

Figure 4 shows the resulting posterior densities; the reference posterior distribution is derived from an improper uniform distribution (e.g. see Modin et al. 2025) and thus equals the data curve in Figure 3.

*<Fig. 4. GREAT example: Probability curves (densities) for 4 posterior distributions; the reference posterior is derived from a uniform prior and thus equal to the data curve in Fig. 3; for ease of substantive interpretation the horizontal axis labels are for RR.>*

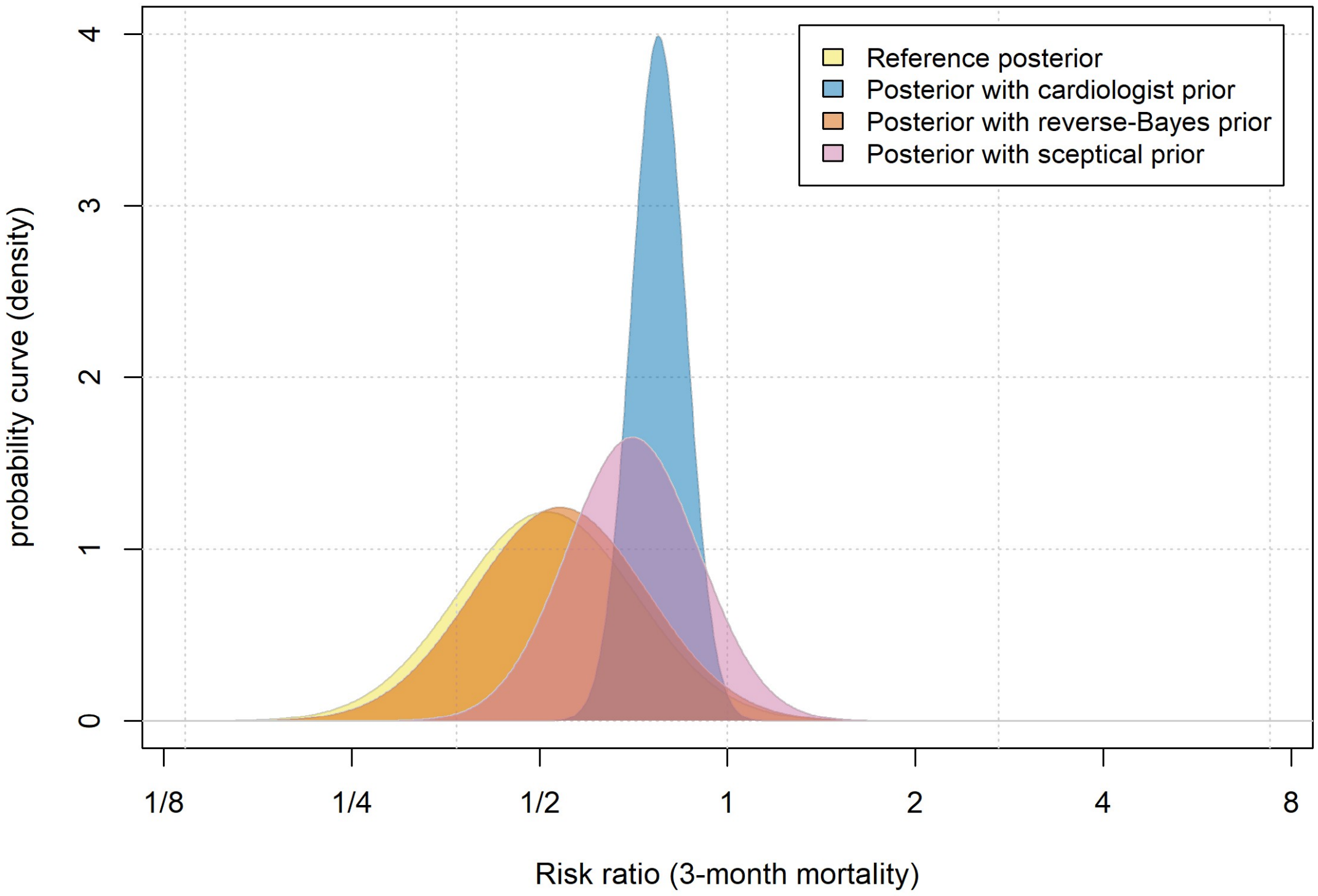

*Comparing prior and data information*

Suppose again that the prior is normal with variance $\sigma^2$ and the frequentist estimator is approximately normal with variance $s^2$, as in the example for the log risk ratio. We can then measure the prior information by $1/\sigma^2$ and the data information by $1/s^2$, and measure the relative contribution of the prior versus the data by the information ratio $(1/\sigma^2)/(1/s^2)$, which equals the variance ratio $s^2/\sigma^2$.

In the GREAT example the variance $s^2$ for the observed log risk ratio is 0.1075, while the variance $\sigma^2$ for the log risk ratio in the expert prior is 0.0170. Hence the information ratio is 0.1075/0.0170 = 6.33, and the prior contributes 0.1075/(0.1075+0.0170) = 86% of the information in the posterior for the risk ratio. This shows the sense in which the posterior in this example is mostly a product of the expert prior, not the actual data. For the equipoise prior with 95% limits of ½, 2 has variance 0.1251, hence the prior contributes 0.1075/(0.1075+0.1251) = 46% of the posterior risk-ratio information. Finally, the reverse-Bayes prior with 95% prior limits of 0.10, 10 has variance 1.381, so this prior contributes 0.1075/(0.1075+1.381) = 7.2% of the posterior information, reflecting that the frequentist upper limit of 0.98 needs little prior input to reach 1.00.

We could also measure and compare effective sample sizes (ESS) contributed by the prior and the data to the Bayesian posterior result. Appendix 2 derives $1+4/\sigma^2$ as a simple intuitive measure of the ESS for a normal prior on a log odds ratio, log hazard ratio, or log risk ratio with mean $\mu$ and variance $\sigma^2$. This measure is found by translating the prior into the result from a thought experiment whose frequentist point estimate and estimated variance match the prior mean and variance (Spiegelhalter et al. 2004, Greenland 2006, Rothman et al. 2008 Ch. 18); generalizations of prior data and hence prior ESS to non-normal priors and to regression coefficients are available.(Greenland 2007, Sullivan and Greenland 2013) In the GREAT example, the expert prior injects information equivalent to a trial with ESS = $1+4/0.0170 \approx 236$ deaths among a very large number of participants in a balanced trial, over 6.5 times the 36 actual deaths in the GREAT trial. The prior with 95% limits of ½, 2 injects information equivalent to $1+4/0.1251 \approx 33$ deaths, not quite as many as actual deaths, while the prior with 95% limits of 0.10, 10 injects the equivalent of only about $1+4/1.381 \approx 4$ deaths.[9]

*Special cautions for sensitivity and bias analyses*

Our example is highly simplified insofar as GREAT was a simple randomized trial about which no impactful methodologic concerns or modeling issues were raised. Observational studies, as well as trials with problems such as extensive nonadherence or drop-out, raise further issues that require careful analyses.

In sensitivity analyses, specifics of the analysis approach such as the underlying causal and measurement models are varied to see how inferences change across model candidates; in bias analyses the specifics are models for bias sources and their effects (Rothman et al. 2008, Ch. 19, Fox et al. 2021, Greenland 2009a, Greenland 2024). Using both frequentist and Bayesian analyses could be seen as part of this variation, but the distinction between them diminishes when, under realistic assumptions, the available data (both current and prior) contain little information about the targeted effect: In that case, the only inference available from competent, honest analyses will be that more research is needed if one wants a more decisive result.

A major problem for all approaches is that the variety of potential candidate models is far too vast to encompass. As a result, sensitivity analyses can mislead via biased or restrictive choices for the candidates, or by including on equal footing candidates that are plausible and candidates that are absurd. To address that problem, one may assign prior distributions across candidate-model parameters based on plausibility considerations, and produce summary results across the candidate analyses (Rothman et al. 2008, Ch. 19, Fox et al. 2021, Greenland 2009a, Greenland 2024). For a Bayesian these summaries are posterior distributions, whereas for frequentists they can be intervals derived from probabilistic sensitivity analysis (Rothman et al. 2008, Ch. 19, Fox et al. 2021) or the analogs of Bayesian results derived by penalization or similar hierarchical, multilevel, or random-coefficient methods (Greenland 2009a, Greenland 2024). In all these cases the results will be completely dependent on the choice of priors, and so all the concerns we have raised apply with maximum force.

---

[9] The approximation used by the formula becomes inadequate when its resulting ESS is 3 or less; Appendix 2 provides a more accurate approach for such cases.

## Discussion

### *The assumptions behind frequentist and Bayesian analyses*

A conventional frequentist analysis of a trial uses information extracted from the analysed data based on assumptions about the mechanisms that generated those data, such as assumptions about treatment randomization, masking of patients and outcome evaluators, causes of drop-out, and the shape of dose-response. Because these assumptions can strongly determine inferences and decisions derived from data analyses, they should be explicated and justified in both a pre-specified analysis plan and in study reports. Problems arising from neglect of explicit justification include use of analysis models that poorly reflect both the actual data-generation process and the final data. A common example is the use of the randomization assumption to estimate effects from "as-treated" and per-protocol analyses of trials with considerable nonadherence and drop-out.

Typical Bayesian analyses employ the same assumptions and models of data generation that are used in frequentist analyses, hence are subject to the same problems from erroneous assumptions and poor data models. They then add a further assumption in the form of a prior distribution for the actual treatment effect size in the trial. The prior can strongly determine inferences and decisions derived from data analyses, and so needs to be explicated and justified in a pre-specified analysis plan and in study reports. This need expands the labor and vulnerabilities of informative-Bayesian analyses beyond those of conventional frequentist analyses.

Bayesian analyses may add further prior distributions for effects of baseline covariates on endpoints, and for effects of treatment and of baseline covariates on competing risks, adherence, and drop-out (Greenland 2005, Gustafson 2005, 2015), although the same expansion can be done using parallel frequentist methods (Greenland 2009a). As with priors for the treatment effect on trial endpoints, depending on their own accuracy these additional priors may help or harm accuracy of the final treatment-effect estimates. Regardless, all priors need to be precisely described and justified in study reports so that the reader may judge their realism and plausibility.

*How Bayesian methods should be integrated into basic statistical training*

Basic Bayesian textbooks have long been available (e.g., Blackwell 1969) but have not dislodged frequentist treatments. Nonetheless, there is a growing sentiment that Bayesian methods are an established part of applied statistics and thus should be taught as part of basic statistics. We agree − as long as that coverage includes ample instruction on how to construct good priors, how to recognize bad ones, and how to vary priors to gauge their impact. We would avoid elementary treatments more heavily focused on mathematics and techniques for producing posterior inferences instead of on practical concerns in prior construction. We have thus attempted to provide a nontechnical overview of what we see as key issues all reviewers and readers of Bayesian analyses should be aware of, and some simple tools to address those issues. That overview has included the foundational differences and parallels between conventional frequentist and Bayesian methodology.

Proponents of exclusively Bayesian approaches argue that those can reduce the proliferation of misleading research reports in ways that cannot be achieved with frequentist methods, and thus should become the norm. Such claims need to go beyond philosophical, mathematical, and semantic arguments to empirical studies. Supporting experiments exist (e.g., Kubsch et al. 2021), and there are observational studies indicating that Bayesian methods can aid in interpreting medical research. As an example, an analysis of studies in the Cochrane collaboration showed how to empirically derive a prior distribution that appeared effective in adjusting for the well-known tendency of early reports to display exaggerated effect estimates. (van Zwet et al, 2021) And an analysis of 230 trials reported that the use of Bayesian methods helped shift focus away from merely testing the null hypothesis of no effect toward evaluating evidence for clinically meaningful effects. (Sherry et al, 2026) The latter shift can be achieved within the frequentist framework in many ways (for example via use of noninferiority and superiority P-values), but this shift has been slow and so there is a hope that Bayesian methods will move it along.

Even if basic coverage of Bayes becomes the norm in elementary statistics, we suspect misuses are inevitable and will at times get by reviewers. A core problem is prior construction and criticism, which is an involved and controversial topic area covered in detail by the exclusively Bayesian statistics literature (e.g., Spiegelhalter et al. 1994; Spiegelhalter et al. 2004 Ch. 5; Gelman et al. 2013; Gelman et al. 2017; van de Schoot et al. 2021; Held 2023). We have found however that this coverage is well beyond the level of typical clinical researchers and readers. It may thus be no surprise that many Bayesian analyses of trials provide no justification and sometimes not even a clear description of the priors in the main study report (Ferreira et al. 2020). We thus argue that Bayesian analyses should be treated as extensions or supplements to conventional analyses, rather than replace them, for that treatment puts prior construction and justification at the fore of the transition from frequentist to Bayesian analyses in teaching and practice.

*Our requirements for Bayesian reporting*

To summarize our position for practice, we would first and foremost require that reports of Bayesian analyses give the priors and their justifications in detail in the main paper (e.g., as in Modin et al., 2025). We shouldn't have to struggle to find the priors buried in an appendix or supplement, as they are often the core determinants of the Bayesian results.

As illustrated in the GREAT example, an expert prior can contribute far more information to the posterior than does the trial data, so that any inference from that posterior is more strongly influenced by the expert's opinion than by the actual data. Furthermore, we would **not** accept undemonstrated claims that a prior is “weakly informative”, as the label encompasses priors that might critically affect borderline results: A prior that seems widely dispersed or neutral, and thus weak to some experts, may nonetheless exert substantial influence on the posterior if the data information fails to overwhelm the prior information about borderline effects.

Thus, following the good practices advised by Spiegelhalter et al. (2004, Ch. 5), our next demand is to see full details of the derivation of the priors, their impact on the posterior, and a comparison of the information contributed by the priors and the data. Specifically:

a) The rationales for the priors should be explained, preferably as part of a publicly registered pre-trial protocol. Priors fail this requirement if they contain more than negligible information and have no reasonable derivation from actual external information or reverse-Bayes analysis.

b) Reports should show the impact of a prior by comparing its prior interval, the posterior interval, and the conventional frequentist interval or reference posterior interval derived from the same data model.
c) Reports should also give direct measures of the relative information taken from the prior and data, such as variance ratios and effective sample sizes. Space permitting, we would also want to see graphical overlays of prior and posterior distributions along with a reference curve showing information from the data alone (e.g., a likelihood function or reference posterior).

We fear that without strong enforcement of these reporting requirements, Bayesian methods will be reduced to a form of circular reasoning that injects confirmation bias into inferences, or worse, serve as a tool for producing desired conclusions from technically rigorous analyses. For regulatory applications, even more extensive guidelines are needed such as those proposed by the U.S. FDA (2026).

**Appendix 1. Frequentist statistics vs. Bayesian statistics**

All statistical analyses start with a model that summarizes assumptions about how the data were generated, then use this data model to extract information from the observed data. For example, simple randomization warrants use of a model that assumes all patients entering the trial have the same chance being assigned to any given treatment group. If there are no losses or other sources of missing data, a frequentist analysis can be based solely on this assumption and the trial data, as in simple permutation statistics. Extensions to deal with problems such as censoring are then made by assuming those events are (like treatment assignment) also random, perhaps within levels of measured covariates. The resulting analyses are thoroughly automated and have been an accepted research standard for going on a century.

That simplicity and acceptance comes at a high cost, however: Almost all conventional frequentist analyses summarize data information in the form of P-values and related interval estimates called “confidence” intervals (CI). Correct interpretations of these P-values and CIs are very subtle and often misunderstood by researchers (Greenland et al. 2016a). Confusion is enhanced by the traditional use of terms like “significance” and “confidence” to label these statistics, because in ordinary English a phrase like “95% confidence” implies a degree of belief that at best applies only under ideal conditions. Thus, based on historical precedents, there is a movement to rename “significance levels” as *compatibility levels*, and rename “confidence intervals” and “credible intervals” as *compatibility intervals* (still CI) to avoid assigning them unwarranted confidence or credibility (Amrhein & Greenland 2022, Berner &

Amrhein 2022, Greenland et al. 2022, Greenland 2025, McShane et al. 2024, Rafi & Greenland 2020, Rovetta et al. 2025ab).

As an example, a conventional risk ratio (RR) estimate of 1.41 with p=0.13 for the null hypothesis that RR=1 and a 95% CI of (0.90,2.20) would often be misinterpreted as supporting the null hypothesis because p>0.05 or because the null value (RR=1) is inside the interval. That is however a serious mistake based on confusing **failure to reject** RR=1 at the 0.05 level with **support** for the null hypothesis (a "nullistic" fallacy; see Greenland 2017, Rafi & Greenland 2020, Greenland et al. 2022). Another common mistake is to say the analysis produced a 95% probability that the true RR is between 0.90 and 2.20; but in frequentist statistics we can only say that (according to our data model) all RR from 0.90 to 2.20 have 2-sided P-values above 0.05 and are thus reasonably compatible with the data by that criterion.

Besides tendencies to be misinterpreted, conventional analyses ignore important background (contextual) information about the likely sizes of effects. For example, by the time of a phase-III trial it is clear that the treatment effect is not dramatic enough to obviate the trial, but this fact is usually ignored in the subsequent analyses. Various methods allow the introduction of such background information into frequentist analyses (e.g., via *penalization* (Greenland 2009a, Cole et al. 2014)). Nonetheless, the most common way to use that information is via *informative-Bayesian* analysis, in which background information is used to develop a *prior probability distribution* (prior) for effect sizes.

In one interpretation of Bayesian statistics (de Finetti 2017), these priors are thought of as supplying bets about the effect sizes before seeing the study data. Thus, a prior distribution with median at a risk ratio of 1 (RR=1) places equal odds on the effect being a preventive (RR<1) vs. a cause (RR>1) of the outcome. If 90% of that prior was below 2 and 10% was above 2, it would be placing 9:1 odds on RR<2 vs. RR>2. These odds are examples of the betting information about RR contained in the prior distribution.

In basic Bayesian analyses, the information in the data about effects is first summarized as probabilities of the data under specific values for the effect and the data model, called *likelihoods*. Ratios of those likelihoods can then be computed; in our example, the likelihood for one value $RR_2$ (e.g., RR=2) can be divided by the likelihood for another value $RR_1$ (e.g., RR=1) to produce the likelihood ratio $LR(RR_2{:}RR_1)$. This summary data information is then merged with the prior information to produce a post-data or *posterior* distribution, which supplies a new set of bets about effect sizes. The merging is done via Bayes theorem, in which the new odds of $RR_2$ versus $RR_1$ can be computed via the formula

$$\text{Posterior odds}(RR_2{:}RR_1) = \text{Prior odds}(RR_2{:}RR_1)\times LR(RR_2{:}RR_1).$$

(Speigelhalter et al. 2004 eq. 3.2, Gelman et al. 2013, McElreath 2020). The posterior odds are said to update the prior odds using the data information, although they can also be interpreted as the odds obtained upon augmenting the data information by the prior information (Greenland 2006; Greenland 2007; Rothman et al. 2008 Ch. 18; Sullivan and Greenland 2013).

If the prior odds are always close to 1, we say the prior information is negligible relative to the data information and that the prior is "nearly noninformative". In that case the posterior odds will always be close to the likelihood ratio. In typical cases this will result in the posterior intervals being *numerically* close to the frequentist CI of the same percentage. Nonetheless, the meaning of the Bayesian and frequentist percentages will differ; for example, a Bayesian 95% posterior interval for the RR of (0.90,2.20) will be interpretable as a bet of 95% (or 19:1 odds) on the hypothesis that RR is between 0.90 and 2.20. In contrast, a frequentist 95% CI of (0.90,2.20) will not have that interpretation; instead it will only display the range of RR that have two-sided P-values above $1-0.95 = 0.05$.

On the other hand, if the amount of prior information is considerable, the posterior interval will not resemble the CI because the latter uses only the data information. This leads us to consider how we might measure and compare the information in the prior and the data. As discussed in Appendix 2, in typical applications a useful information measure is the inverse variance, which can be translated into a measure of *effective sample size* (ESS).

**Appendix 2. A simple measure of effective sample size**

There are many ways to define and measure effective sample size, several of which are cited by the US FDA (2026, p. 22). We describe one that translates the prior into data from a randomized trial with a binary outcome (e.g. death/no death) and no censoring, as shown in a 2x2 table with cells a = number of deaths in a treatment arm, b = number of deaths in a control or comparator arm and c and d are the numbers of non-deaths in the respective arms of the trial:

| | Treated | Control | |
|---|---|---|---|
| Death | a | b | $M_1 = a+b$ |
| Survival | c | d | $M_0 = c+d$ |
| Allocation | $N_1 = a+c$ | $N_0 = b+d$ | |

Two common estimates of effect from such data are the odds ratio OR = (a/c)/(b/d) = (a*d)/(b*c) and the risk ratio RR = $(a/N_1)/(b/N_0)$. The ratio HR of average hazard rates is also commonly used, but requires the survival times to compute so we omit it here; it will fall between the risk ratio and the odds ratio (Rothman et al. 2008, Ch. 4). We will show how to build a simple table whose usual frequentist point and variance estimate for the natural log of the risk ratio, ln(RR), matches the mean $\mu$ and variance $\sigma^2$ of a normal prior for ln(RR). Within that table we will identify $M_1 = a+b$ with the effective sample size (ESS) of the prior distribution, and add a small correction to it.

We start by imagining the normal prior distribution came from a perfect prior trial with no censoring, nonadherence, or other problems, and that yielded data as shown in the table. The usual approximate variance estimate for the log risk ratio from this thought experiment is then $V = 1/a + 1/b - 1/N_1 - 1/N_0$ (Rothman et al. 2008, p. 249) We further imagine this trial had a very low death rate in both arms, so that $N_1$ and $N_0$ are both very large compared to a and b, making $1/N_1$ and $1/N_0$ negligible. The differences between the odds ratio, hazard ratio, and risk ratio will then be negligible, and all will have a variance $V \approx 1/a + 1/b$. This shows that the sizes of a and b are what determine the amount of information about RR in our hypothetical trial. A Bayesian may imagine a and b as the approximate numbers of treated and control deaths expected in $M_1$ draws from the prior predictive distribution (Box 1980, Gelman et al. 2013) for deaths in a randomized trial of $N_1$ treated and $N_0$ controls when the death rates in both arms are very low.

Next, to make the log risk ratio of the table equal to the prior mean $\mu$ we set a=b and note that, with this equality, RR becomes $(a/N_1)/(b/N_0) = N_0/N_1$. We thus set $N_0/N_1$ equal to $e^{\mu}$, the prior median for RR, which is 0.7746 in the GREAT example. This makes $\ln(RR) = \mu$, $M_1 = a+b = 2a$, and $V \approx 2/a = 4/2a = 4/M_1$. Finally, we set this approximate V of $4/M_1$ equal to the prior variance $\sigma^2$, which yields $M_1 = 4/\sigma^2$, as in Spiegelhalter et al. (2004, p. 26). The classical "continuity correction" of adding 0.5 to all cells to improve the approximation translates to an ESS of $1+M_1$.(Sullivan and Greenland 2013) Because we have imposed very low death rates on the prior table, the ESS will be the same if we use instead the odds ratio or hazard ratio. For the GREAT example with the expert prior these formulas lead to $\mu$= ln(0.7746), $N_0/N_1$= 0.7746, , ESS ≈ $1+M_1$ = 1+4/0.0170 ≈ 236 and thus a=b=118.

The construction used above assumes $M_1$ is large enough to justify the usual approximately normal analysis of the prior data. An approach that bypasses this approximation starts instead from log-F($M_1$,$M_1$) priors, which provide natural conjugate and reference priors for logistic and loglinear regressions, and inject the equivalent of ESS = $M_1$ deaths into those regressions (Greenland 2007, Greenland and Mansournia 2015, Greenland et al. 2016b). These priors approach normality as $M_1$ increases but have increasingly heavy tails as $M_1$ gets small.

As an example, the reference prior with a 95% prior interval of (1/648, 648) mentioned in the text is the log-F(1,1) prior and has an ESS of 1; for log rate ratios it is the Jeffreys prior and injects information equivalent to adding ½ to each death count (Greenland and Mansournia 2015). In contrast, the normal prior from $M_1$=1, a=b=0.5 produces V=4 which yields an approximate 95% interval of exp(±1.96·2) = (1/50, 50) and ESS ≈ 1+4/4 = 2. Similarly, a log-F(2,2) prior imposes a 95% prior interval of (1/39,39) and has an ESS of 2; for log rate ratios it is the logistic reference prior and injects information equivalent to adding 1 to each death count. In contrast, the normal prior from $M_1$=2, a=b=1 produces V=2 which yields an interval of exp(±1.96·sqrt(2)) = (1/16, 16) and ESS ≈ 1+4/2 = 3. More generally, the log-F(2a,2b) family can provide skewed priors when a≠b (Greenland 2007).